\documentclass[a4paper, 10pt, onecolumn]{article}
\pdfoutput=1
\usepackage[left=1.7cm,top=1.8cm,right=1.7cm,bottom=1.8cm,nohead]{geometry}
\begin{document}
\title{\large \bf Reply to the ``Comment on: Testing the speed of `spooky action at a distance' ''}
\author{\normalsize D. Salart, A. Baas, C. Branciard, N. Gisin, and H. Zbinden\\
\it \small Group of Applied Physics, University of Geneva, 20, Rue de l'Ecole de M\'edecine, CH-1211 Geneva 4, Switzerland}
\date{\small \today}
\maketitle

{\small
\begin{center}
This is the reply to the comment arXiv:0810.4452 by Kofler, Ursin, Brukner, and Zeilinger.
\end{center}
}

Quantum correlations cannot be described by local common causes. This prediction of
quantum theory, surprising as it might appear, has been widely confirmed by numerous
experiments. In our Nature Letter \cite{SpookyAction} we considered this point as established and addressed
another issue: the alternative assumption that quantum correlations are due to supra-luminal
influences of a first event onto a second event. For this purpose we believe that it suffices to
observe 2-photon interferences with a visibility high enough to potentially violate Bell's
inequality, as we reported (over 2 x 17.5 km). Simultaneously closing other loopholes, like
the locality loophole as desired by Kofler and colleagues, would certainly be an interesting
addition, as would be any Bell tests that simultaneously address several of the loopholes.
Indeed, to rigorously exclude any common cause explanation of the observed quantum
correlation one should, ideally, simultaneously close the locality and the detection loophole
(and assume the existence of independent randomness and that quantum measurements are
finished when detectors fire – or at least when a mesoscopic mass has sufficiently moved as
insured in our experiment, see our recent article \cite{MovingMass}). This is a
formidable task and any progress towards achieving it is most welcome. So far, however, all
experiments have addressed at most one of these loopholes; ours is no exception.\\

Concerning the comment on the use of a Franson interferometer for testing quantum
nonlocality, we stress that this is not a fundamental issue. In principle it suffices to replace the
entrance beam splitters of each interferometer by a fast switch. In this way the non-interfering
lateral peaks observed in the 2-photon interferogram would disappear. However, in practice
such switches suffer due to losses of around 3 dB. Hence, with today's technology it is much
more convenient to replace the ideal switch by a passive coupler, as we did in our experiment
in a way very similar to \cite{PassiveCoupler}.

\begin{center}
\line(1,0){150}
\end{center}
\end{document}